\documentclass[11pt,a4paper]{article}
\usepackage{jheppub}

\title{NLO QCD corrections to dijet production via quark contact interactions}

\author[a]{Jun Gao}
\author[b]{Chong Sheng Li}
\author[c,d]{and C.-P. Yuan}

\affiliation[a]{Department of Physics, Southern Methodist
University, Dallas, TX 75275-0175, USA} \affiliation[b]{Department
of Physics and State Key Laboratory of Nuclear Physics and
Technology, Peking University, Beijing 100871 China}
\affiliation[c]{Department of Physics and Astronomy, Michigan State
University, East Lansing, 48824, USA} \affiliation[d]{Center for
High Energy Physics, Peking University, Beijing, China}

\emailAdd{jung@smu.edu}
\emailAdd{csli@pku.edu.cn}
\emailAdd{yuan@pa.msu.edu}

\abstract {We present the exact next-to-leading order (NLO) QCD
corrections to dijet production at the LHC via quark contact
interactions, with different color and chiral structures induced
from new physics. Following the recent analysis of quark
compositeness search at the LHC, we find that the NLO QCD
corrections can lower the dijet cross sections by several tens
percent, depending on the theory parameters and the selected
kinematic regions, and reduce the dependence of the cross sections
on factorization and renormalization scales. We also calculate the
renormalization group (RG) improved NLO cross sections by summing
over the large logarithms from the RG running of Wilson
coefficients. Moreover, we investigate the NLO QCD effects on
various experimental observables and exclusion limits of quark
compositeness scale.}


\begin{document}

\begin{flushright}
  SMU-HEP-12-08\\
  20th April 2012
\end{flushright}

\maketitle

\section{Introduction}
Jet production at hadron colliders provides an excellent opportunity
to test perturbative QCD (PQCD) and to search for possible new
physics (NP) beyond the Standard Model (SM) over a wide range of
energy scales. Invariant mass distributions of the
dijets~\cite{Chatrchyan:2011ns}, the dijet angular
distributions~\cite{Khachatryan:2011as,Aad:2011aj,Chatrchyan:2012bf},
and other jet observables at the LHC~\cite{:2010wv} have already
extended current searches for quark compositeness, excited quarks,
and other new particle resonances toward the highest energies
attainable. Among all these measurements, the dijet angular
distribution shows a great sensitivity to possible quark contact
interactions induced by new physics models. In the SM, Quantum
Chromodynamics (QCD) predicts the jets in dijet events are
preferably produced in large rapidity region, via small angle
scattering in t-channel processes. On the contrary, the dijet
angular distribution induced by quark contact interactions is
expected to be much more isotropic, and thus the dijet angular
distribution at the LHC could be largely modified.

The measurement of quark contact interactions has been used to set
limits on the quark composite models which have been studied
extensively in the literature~\cite{Eichten:1983hw,Lane:1996gr}. It
is assumed that quarks are composed of more fundamental particles
with new strong interactions at a compositeness scale $\Lambda$,
much greater than the quark mass scales. At the energy well below
$\Lambda$, quark contact interactions are induced by the underlying
strong dynamics, and yield observable signals at hadron colliders.
The newest bounds of $\Lambda$ at the $95\%$ confidence level (C.L.)
from the CMS collaboration are around $10\,{\rm
TeV}$~\cite{Chatrchyan:2012bf} based on $2.2\,{\rm fb}^{-1}$
collected data. Previous limits from the Tevatron and LHC can be
found in
Refs.~\cite{Khachatryan:2011as,Aad:2011aj,Abe:1996mj,Collaboration:2010eza}.
With more integrated luminosity collected, these limits will be
further improved.

In our previous work~\cite{Gao:2011ha}, we carried out the
next-to-leading order (NLO) QCD correction for dijet production at
the LHC induced by the quark contact interactions that are the
products of left-handed electroweak isoscalar quark currents. We
compared it with the leading order (LO) results used by the CMS
Collaboration and also the ``scaled NLO results'' used by the ATLAS
Collaboration, which assumes the NLO correction (in terms of
K-factors, defined as the ratio of NLO cross sections to LO ones) to
the dijet production from contact interactions to be exactly the
same as that from the SM QCD interactions. And we derived, based on
our exact NLO results, the corrected limits of compositeness scale
for the CMS and ATLAS measurements with $3\,{\rm pb}^{-1}$
data~\cite{Collaboration:2010eza}. In this paper, we extend our
previous work to include more quark contact interaction operators in
the NLO QCD calculations, which allows the mixing of operators with
different chiral structures at the NLO level, and show more details
of the calculations. The effect of our results to measurements of
quark compositeness at the LHC is also investigated. In the
appendix, we discuss some details of
our numerical code developed for the calculations presented in this paper.

\section{Theoretical setup}
We consider a subset of quark contact interactions that are the
products of electroweak isoscalar quark currents which are assumed
to be flavor-symmetric to avoid large flavor-changing
neutral-current interactions~\cite{Lane:1996gr}. The effective
Lagrangian can be written as
\begin{equation}
\mathcal{L}_{NP}=\frac{1}{2\Lambda^{2}}\sum_{i=1}^6c_{i}O_{i},
\end{equation}
where $\Lambda$ is the new physics scale, $c_i$ are Wilson
coefficients. And the operators $O_i$ in chiral basis are given by
\allowdisplaybreaks{\begin{eqnarray} O_{1} & = &
\delta_{ij}\delta_{kl}\left(\sum_{c=1}^{3}\bar{q}_{Lci}\gamma_{\mu}
q_{Lcj}\sum_{d=1}^{3}\bar{q}_{Ldk}\gamma^{\mu}q_{Ldl}\right),\nonumber \\
O_{2} & = & {\rm T}_{ij}^{a}{\rm T}_{kl}^{a}\left(\sum_{c=1}^{3}\bar{q}_{Lci}
\gamma_{\mu}q_{Lcj}\sum_{d=1}^{3}\bar{q}_{Ldk}\gamma^{\mu}q_{Ldl}\right),\nonumber \\
O_{3} & = & \delta_{ij}\delta_{kl}\left(\sum_{c=1}^{3}\bar{q}_{Lci}\gamma_{\mu}
q_{Lcj}\sum_{d=1}^{3}\bar{q}_{Rdk}\gamma^{\mu}q_{Rdl}\right),\nonumber \\
O_{4} & = & {\rm T}_{ij}^{a}{\rm T}_{kl}^{a}\left(\sum_{c=1}^{3}\bar{q}_{Lci}
\gamma_{\mu}q_{Lcj}\sum_{d=1}^{3}\bar{q}_{Rdk}\gamma^{\mu}q_{Rdl}\right),\nonumber\\
O_{5} & = & \delta_{ij}\delta_{kl}\left(\sum_{c=1}^{3}\bar{q}_{Rci}\gamma_{\mu}
q_{Rcj}\sum_{d=1}^{3}\bar{q}_{Rdk}\gamma^{\mu}q_{Rdl}\right),\nonumber \\
O_{6} & = & {\rm T}_{ij}^{a}{\rm
T}_{kl}^{a}\left(\sum_{c=1}^{3}\bar{q}_{Rci}
\gamma_{\mu}q_{Rcj}\sum_{d=1}^{3}\bar{q}_{Rdk}\gamma^{\mu}q_{Rdl}\right),
\end{eqnarray}}in which $c$, $d$ are generation indices and $i$, $j$, $k$, $l$, $a$
are color indices, and ${\rm T}^{a}$ are the Gell-Mann matrices with
the normalization ${\rm Tr}({\rm T}^{a}{\rm T}^b)=\delta^{ab}/2$.
Beside of the quark compositeness, the above interactions can also
arise from various kinds of new physics models, induced by the
exchange of new heavy resonances, such as $Z'$
models~\cite{Langacker:2008yv} and extra dimensions
models~\cite{Randall:1999ee}. Thus our analyses here are rather
model independent and $\Lambda$ can be identified as the effective
new physics (NP) scale.

The above six operators have been extensively studied in weak decays
of mesons~\cite{Buchalla:1995vs}. They mix with each other through
QCD loop diagrams, which requires a renormalization matrix of the
operators to cancel all the ultraviolet divergences, defined by
$O_i^{(0)}=(1+\delta Z)_{ij}O_{j}$. After calculating one-loop
diagrams, to be shown latter, with the dimensional regularization scheme in
$n=4-2\epsilon$ dimensions, we obtain the matrix at the NLO as follow
\begin{equation}\label{renor}
\delta Z=D_{\epsilon}{\alpha_s\over 4\pi}{1\over \epsilon}\left(
        \begin{array}{cccccc}
          0 & -{22\over 3} &0&-{4\over 3}&0&0\\
          -{3C_F\over N} & {20\over 3N}-{2\over 3}n_f &0&{2\over 3N}-{2\over 3}n_f&0&0\\
          0 & 0 &0&6&0&0\\
          0 & -{n_f\over 3} &{3C_F\over N}&6C_F-{3\over N}-{2\over 3}n_f&0&-{n_f\over 3}\\
          0 & 0 &0&-{4\over 3}&0&-{22\over 3}\\
          0&0&0&{2\over 3N}-{2\over 3}n_f&-{3C_F\over N} & {20\over 3N}-{2\over 3}n_f\\
        \end{array}
      \right),
\end{equation} where$D_{\epsilon}=\frac{(4\pi)^{\epsilon}}{\Gamma(1-\epsilon)}$,
$N=3$ and $C_{F}=4/3$ for QCD, and $n_f=5$ is the activate quark
numbers in the loop. (Here, we do not include the top quark contribution
in the loops).

\section{Analytical results}

\subsection{NLO corrections}\label{asec1}
At LO, there are several subprocesses which contribute to the dijet
production at hadron colliders induced by the operators
under consideration. They are
\begin{equation}
qq'(q)\rightarrow qq'(q),\ q\bar{q}'\rightarrow q\bar{q}',\
q\bar{q}\rightarrow q\bar{q}(q'\bar{q}'),
\end{equation}where $q$, $q'$ could be all the light quarks except the top quark.
The NP contributions included in our calculation consist of two
parts, the NP squared terms and the interference terms between the
NP and the SM QCD interactions, which have different behavior with
the increase of dijet invariant mass. We carried out the NLO
calculations in the Feynman-'t Hooft gauge with dimensional
regularization (DR) scheme (with naive $\gamma_5$
prescription)~\cite{Buchalla:1995vs} in $n=4-2\epsilon$ dimensions
to regularize all the divergences. Below, we only show the
analytical results for the subprocess $q(p_{1})q'(p_{2})\rightarrow
q(p_{3})q'(p_{4})$, since the similar results for other subprocesses
can be obtained by crossing symmetry.
\begin{figure}[h]\centering
\includegraphics[width=0.5\textwidth]{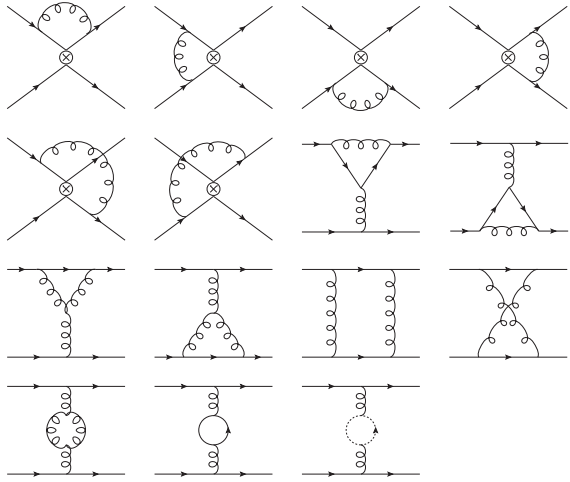}
\caption[]{Loop diagrams for both the SM QCD and NP contributions
to $q q' \to q q'$ at the NLO.} \label{loop1}
\end{figure}

\begin{figure}[h]\centering
\includegraphics[width=0.5\textwidth]{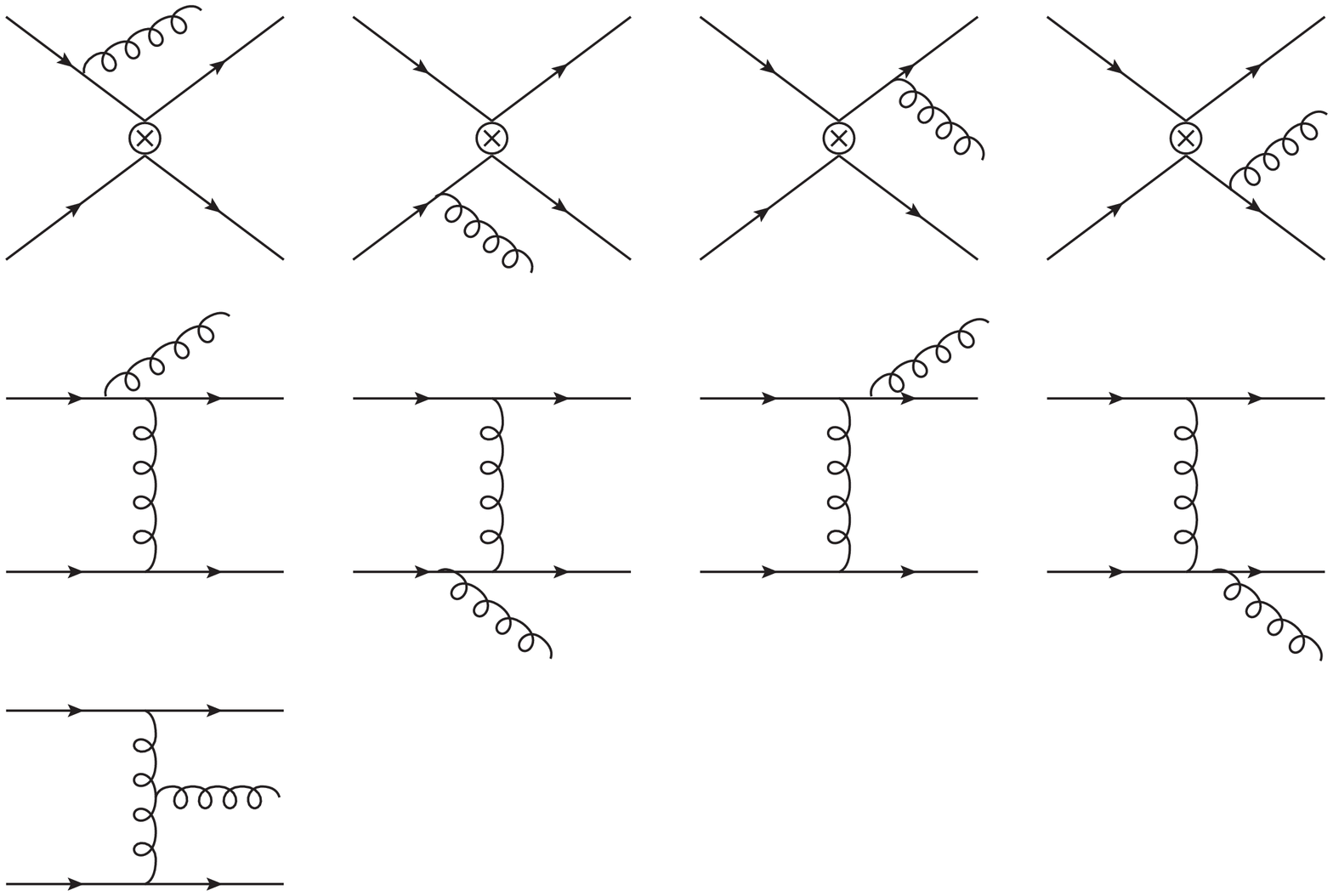}
\caption[]{Real emission diagrams for both the SM QCD and NP contributions
to $q q' \to q q'$  at the NLO.} \label{real}
\end{figure}

First, we define the following abbreviations for color structures
and matrix elements, \allowdisplaybreaks{\begin{eqnarray}
\mathcal{M}_{1} & = & \bar{u}_{L}(p_{3})\gamma_{\mu}u_{L}(p_{1})
\bar{u}_{L}(p_{4})\gamma^{\mu}u_{L}(p_{2}),\nonumber \\
\mathcal{M}_{2} & = & \bar{u}_{R}(p_{3})\gamma_{\mu}u_{R}(p_{1})
\bar{u}_{R}(p_{4})\gamma^{\mu}u_{R}(p_{2}),\nonumber \\
\mathcal{M}_{3} & = & (\bar{u}_{L}(p_{3})\gamma_{\mu}u_{L}(p_{1})
\bar{u}_{R}(p_{4})\gamma^{\mu}u_{R}(p_{2})\nonumber\\
&&+\bar{u}_{R}(p_{3})\gamma_{\mu}u_{R}(p_{1})
\bar{u}_{L}(p_{4})\gamma^{\mu}u_{L}(p_{2}))/2,\nonumber \\
\mathcal{C}_{1} & = & \delta_{i_{3}i_{1}}\delta_{i_{4}i_{2}},\
\mathcal{C}_{2}={\rm T}_{i_{3}i_{1}}^{a}{\rm T}_{i_{4}i_{2}}^{a},
\end{eqnarray}}where $i_{1-4}$ are the color indices of the external quarks. The LO
scattering amplitudes induced by the NP and the SM QCD interactions
can be separately written as
\allowdisplaybreaks{\begin{eqnarray}\label{eqtree}
i\mathcal{M}_{NP,1}^{tree} & = &
i\mathcal{M}_{1}(c_{1}\mathcal{C}_{1}
+c_{2}\mathcal{C}_{2})/\Lambda^{2},\nonumber \\
i\mathcal{M}_{NP,2}^{tree} & = &
i\mathcal{M}_{2}(c_{5}\mathcal{C}_{1}
+c_{6}\mathcal{C}_{2})/\Lambda^{2},\nonumber \\
i\mathcal{M}_{NP,3}^{tree} & = &
i\mathcal{M}_{3}(c_{3}\mathcal{C}_{1}
+c_{4}\mathcal{C}_{2})/\Lambda^{2},\nonumber \\
i\mathcal{M}_{SM,1}^{tree} & = &
i\mathcal{M}_{1}(4\pi\alpha_{s}\mathcal{C}_{2})/t,\nonumber\\
i\mathcal{M}_{SM,2}^{tree} & = &
i\mathcal{M}_{2}(4\pi\alpha_{s}\mathcal{C}_{2})/t,\nonumber\\
i\mathcal{M}_{SM,3}^{tree} & = &
i\mathcal{M}_{3}(8\pi\alpha_{s}\mathcal{C}_{2})/t,
\end{eqnarray}}where $s,\ t,\ u$ are the Mandelstam variables, and we divide both
the NP and SM QCD amplitudes into 3 groups for convenience. After
adding the 1-loop amplitudes, as shown in Fig.~\ref{loop1}, and the
counterterms from renormalization, we obtain  the ultraviolet finite
virtual amplitudes as follows. \allowdisplaybreaks{\begin{eqnarray}
i\mathcal{M}^{v,
uv}_{NP,1}&=&i\mathcal{M}_1C_{\epsilon}{\alpha_s\over 4\pi}
\Big\{-2C_F\Big[c_1\mathcal{A}
(t)+{c_2\over 2N}\mathcal{B}(u)\Big]\mathcal{C}_1\nonumber\\
&&\hspace{-1cm}+\Big[ c_1\big(-2\mathcal{B}(u)\big)+c_2\big(-2C_F
\mathcal{A}(u)+{1\over N}\big(\mathcal{B}(u)+\nonumber\\
&&\hspace{-1cm} \mathcal{B}(t)
\big)\big)\Big]\mathcal{C}_2\Big\}/\Lambda^2,\nonumber\\
i\mathcal{M}^{v, uv}_{NP,2}&=&i\mathcal{M}^{v,
uv}_{NP,1}\{\mathcal{M}_1\rightarrow\mathcal{M}_2, c_1\rightarrow c_5, c_2\rightarrow c_6\},\nonumber\\
i\mathcal{M}^{v, uv}_{NP,3}&=&i\mathcal{M}^{v,
uv}_{NP,1}\{\mathcal{M}_1\rightarrow\mathcal{M}_3, c_1\rightarrow c_3, c_2\rightarrow c_4\}\nonumber\\
&&\hspace{-1cm}+i\mathcal{M}_3{\alpha_s\over
4\pi}\Big\{2C_F\Big[{c_4\over
2N}\big(\mathcal{S}(s)+\mathcal{S}(u)\big)\Big]
\mathcal{C}_1+\Big[2c_3\big(\mathcal{S}(s)\nonumber\\
&&\hspace{-1cm}+\mathcal{S}(u)\big)+c_4\big(2C_F\mathcal{S}(u)-{1\over
N}\big(2\mathcal{S}(s)
+\mathcal{S}(u)\big)\big)\Big]\mathcal{C}_2\Big\}/\Lambda^2,\nonumber\\
i\mathcal{M}^v_{SM,1}&=&i\mathcal{M}_1C_{\epsilon}{\alpha_s\over
4\pi} \Big\{4\pi\alpha_s\Big[-{C_F\over 2N}\Big({4\over
\epsilon}\ln(-{s\over u})-
\nonumber\\
&&\hspace{-1cm}2{t\over s}\ln({t\over u})-{u^2\over s^2}\ln^2({t\over u})+\ln^2({s^2
\over t u})+(1-{u^2\over
s^2})\pi^2\Big)\Big]\nonumber\\
&&\hspace{-1cm}\mathcal{C}_1+4\pi\alpha_s\Big[-2C_F\Big( {2\over
\epsilon^2}+{1\over\epsilon}\big(3+2\ln(-{s\over u})\big)\Big)\nonumber\\
&&\hspace{-1cm}+{2\over N\epsilon}\ln({s^2\over t u})
+\beta_0\ln({\mu_R^2\over s})-\big({2\over3}n_f-{10\over3}C_F-{8\over3N}\big)\nonumber\\
&&\hspace{-1cm}\ln(-{s\over t})+{3\over N}\ln^2(-{s\over t}) -\big({1\over
2N}-C_F\big)\Big({u^2\over s^2}\big(\ln^2({t \over
u})\nonumber\\
&&\hspace{-1cm}+\pi^2\big)-2{u\over s}\ln({t\over u})+\ln^2({t\over u})-
2\ln(-{s\over u})\big(1\nonumber+\\
&&\hspace{-1cm}\ln(-{s\over u})\big)\Big)
+\big(C_F+{3\over2N}\big)\pi^2-\big({10\over9}n_f-{26\over9}C_F\nonumber\\
&&\hspace{-1cm}-{85\over9N}\big) \Big]\mathcal{C}_2\Big\}/t,
\nonumber\\
i\mathcal{M}^{v}_{SM,2}&=&i\mathcal{M}^{v}_{SM,1}\{\mathcal{M}_1
\rightarrow\mathcal{M}_2\}, \nonumber\\
i\mathcal{M}^{v}_{SM,3}&=&i\mathcal{M}^{v}_{SM,1}\{\mathcal{M}_1
\rightarrow2\mathcal{M}_3\}\nonumber\\
&&\hspace{-1cm}+2i\mathcal{M}_3{\alpha_s\over
4\pi}\Big\{4\pi\alpha_s\Big[{C_F\over
2N}\big(\mathcal{Q}(s/t)+\mathcal{Q}(u/t)\big)\Big]
\mathcal{C}_1\nonumber\\
&&\hspace{-1cm}+4\pi\alpha_s\Big[-{1\over N}
\mathcal{Q}(s/t)+\big(C_F-{1\over
2N}\big)\mathcal{Q}(u/t)\Big]\mathcal{C}_2\Big\}/t, \label{e1}
\end{eqnarray}}where $C_{\epsilon}=({4\pi\mu_R^2\over
s})^{\epsilon}{1\over\Gamma(1-\epsilon)}$, and
\allowdisplaybreaks{\begin{eqnarray} \mathcal{B}(x)&=&{2\over
\epsilon}\ln(-{s\over x})+3\ln(-{\mu_R^2\over x})
+\ln^2(-{s\over x})+\pi^2+9,\nonumber\\
\mathcal{A}(x)&=&{2\over \epsilon^2}+{3\over \epsilon}+({2\over
\epsilon}+3)\ln(-{s\over x})+\ln^2(-{s\over x})+8,\nonumber\\
\mathcal{Q}(x)&=&-{2\over 1+x}\ln(-x)+{1+2x\over
(1+x)^2}\ln^2(-x),\nonumber\\
\mathcal{S}(x)&=&3\big(\ln(-{\mu_R^2\over x})+2\big).
\end{eqnarray}}The superscript ${uv}$ in $\mathcal{M}^{v}_{NP}$ is to indicate that
they are ultraviolet finite.
We have checked that the virtual correction for the SM QCD
contributions given in Eq.~(\ref{e1}) agrees with the ones shown in
Ref.~\cite{Ellis:1985er}. The infrared divergences in virtual
corrections should cancel with those in real corrections. As for the
real corrections, we apply both the phase space slicing based two
cutoff method~\cite{Harris:2001sx} and the subtraction based dipole
method~\cite{Catani:1996vz} in our calculations for a cross-check.
The real emission diagrams include those shown in Fig.~\ref{real} and
all their crossing diagrams.

\begin{figure}[h]\centering
\includegraphics[width=0.6\textwidth]{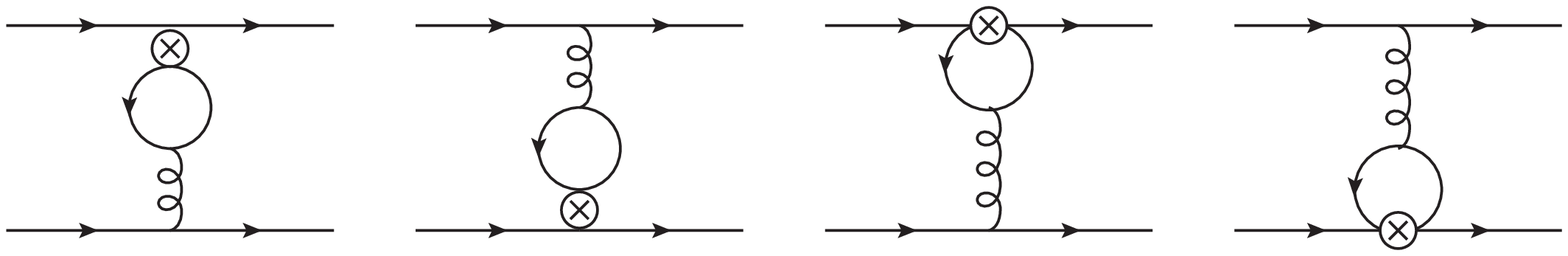}
\caption[]{Additional penguin-like loop diagrams contributed to $q q' \to q q'$
 at the NLO.} \label{loop2}
\end{figure}
For the new physics contributions, beside from the loop diagrams in
Fig.~\ref{loop1}, there are some additional penguin-like loop
diagrams as shown in Fig.~\ref{loop2}. They generate infrared finite
terms with mixing among operators with different chiral
structures. After renormalization, we obtain both infrared and
ultraviolet finite amplitudes \allowdisplaybreaks{\begin{eqnarray}
i\mathcal{M}^{v}_{NP,1}&=&
i(\mathcal{M}_1+\mathcal{M}_3){\alpha_s\over4\pi}\Big\{\Big[-{4\over
3}(c_1-{c_2\over 2N}) \big(\ln({\mu_R^2\over
-t})\nonumber\\
&&+{2\over 3}\big)-{2n_f\over 3}c_2\big(\ln({\mu_R^2\over -t})
+{5\over 3}\big)\Big]\mathcal{C}_2\Big\}/\Lambda^2,\nonumber\\
i\mathcal{M}^{v}_{NP,2}&=&i\mathcal{M}^{v}_{NP,1}
\{\mathcal{M}_1\rightarrow\mathcal{M}_2, c_1\rightarrow c_5,
c_2\rightarrow c_6\},\nonumber\\
i\mathcal{M}^{v}_{NP,3}&=&
i(\mathcal{M}_1+\mathcal{M}_2+2\mathcal{M}_3){\alpha_s\over4\pi}\Big\{\Big[-{n_f\over
3}c_4\big(\ln({\mu_R^2\over -t})\nonumber\\
&&+{5\over 3}\big)\Big]\mathcal{C}_2\Big\}/\Lambda^2,
\end{eqnarray}}which will contribute to the NLO results through interference with
$i\mathcal{M}_{NP}^{tree}$ and $i\mathcal{M}_{SM}^{tree}$ in
Eq.~(\ref{eqtree}).

\subsection{Renormalization group running of Wilson
coefficients}\label{rgc} 
\begin{figure}[h]\centering
\includegraphics[width=0.35\textwidth]{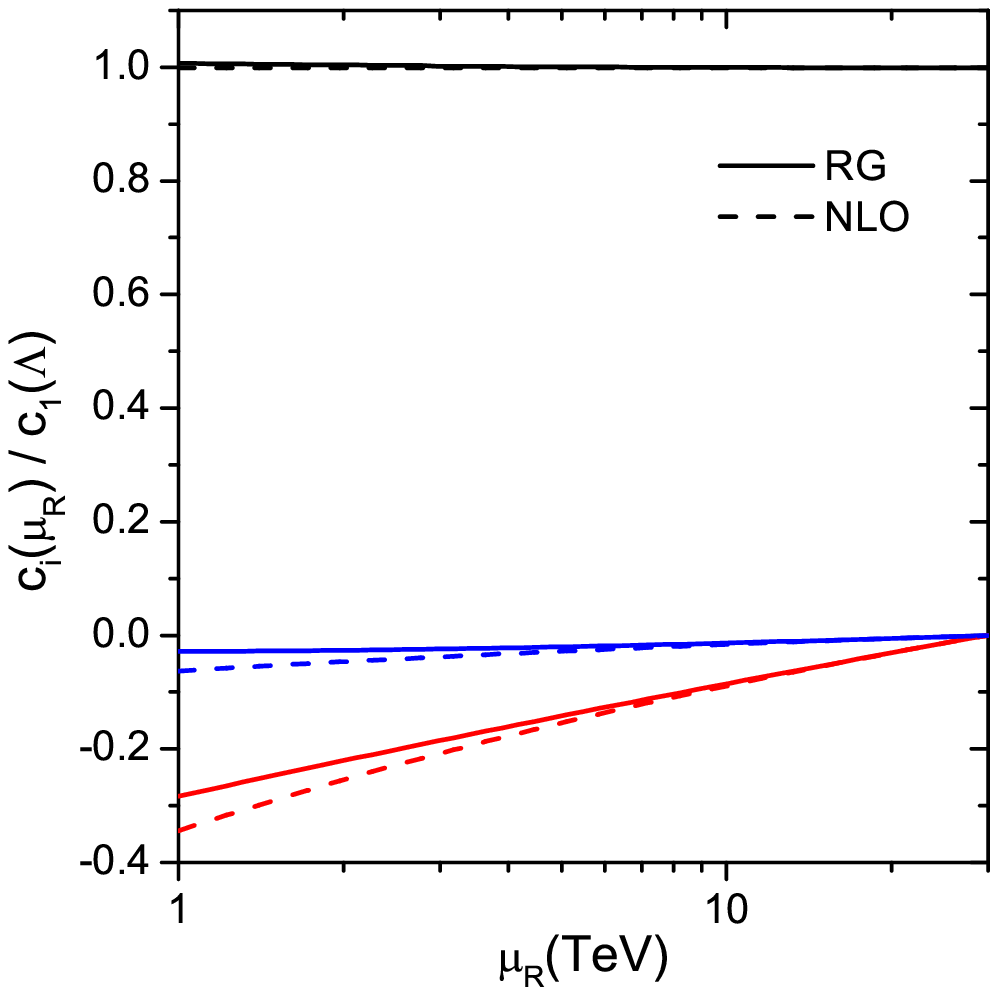}
\hspace{0.2in}
\includegraphics[width=0.35\textwidth]{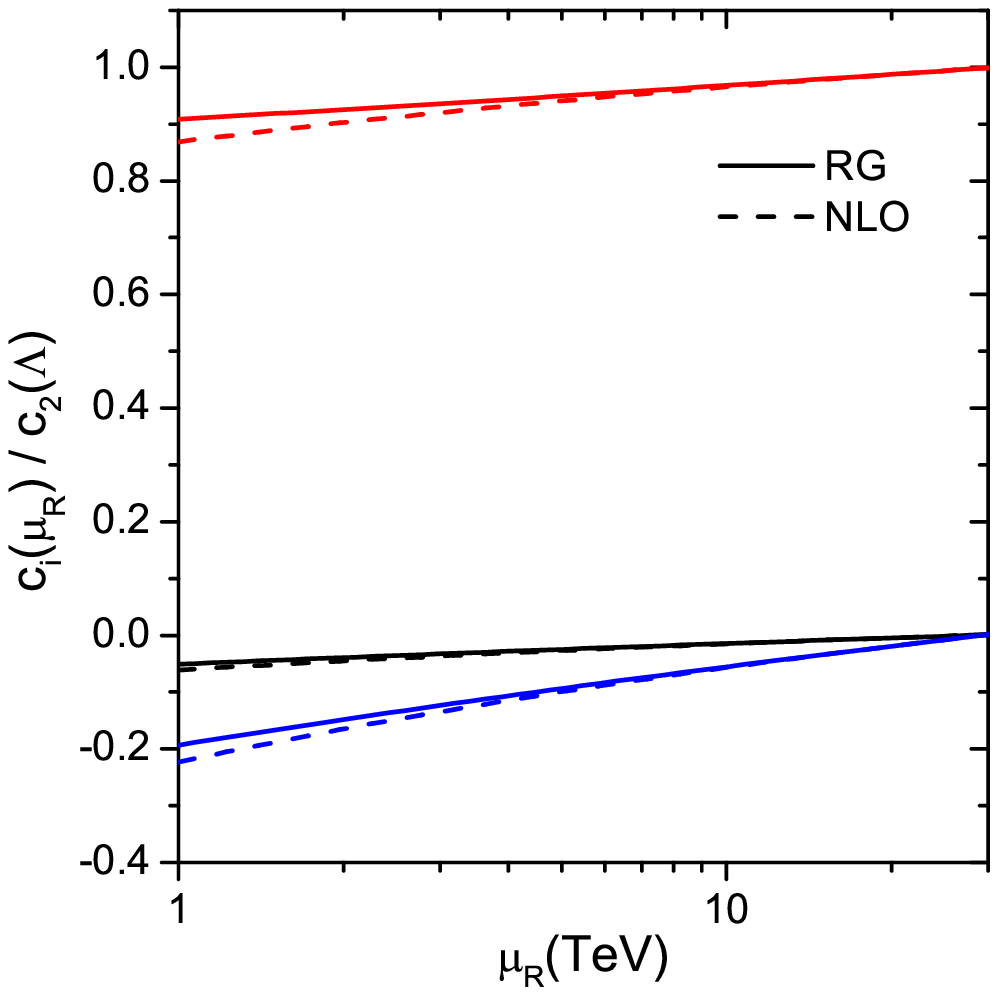}
\caption[]{NLO and RG improved running of the Wilson coefficients.
Curves from top to bottom correspond to $c_1$ (black), $c_4$ (blue)
and $c_2$ (red) for the left panel, and $c_2$ (red), $c_1$ (black)
and $c_4$ (blue) for the right panel.} \label{f2}
\end{figure}

If the new physics scale $\Lambda$ at which
the Wilson coefficients are derived is much higher than the physics
scale considered at colliders, there will be large logarithm terms
associated with these two scales in fixed order calculations. We can
improve the convergence of the perturbative calculation by summing
the logarithm contributions using the renormalization group (RG)
evolution of the Wilson coefficients. The RG equation is given by
\begin{equation}
{d\ c_i(\mu_R)\over d\ \ln\mu_R}=\gamma(g)_{ij}\ c_j(\mu_R),
\end{equation}with the one-loop anomalous dimension matrix derived from
Eq.~(\ref{renor}),
\begin{equation}
\gamma(g)=-{\alpha_s\over 2\pi}\left(
        \begin{array}{cccccc}
          0 &-{3C_F\over N} &0 &0 &0 &0\\
          -{22\over 3} & {20\over 3N}-{2\over 3}n_f &0 &-{n_f\over 3} &0 &0\\
          0&0&0&{3C_F\over N}&0&0\\
          -{4\over 3}&{2\over 3N}-{2\over 3}n_f&6&6C_F-{3\over N}-{2\over 3}n_f& -{4\over 3}&{2\over 3N}-{2\over 3}n_f\\
          0&0&0&0&0&-{3C_F\over N}\\
          0&0&0&-{n_f\over 3}&-{22\over 3}&{20\over 3N}-{2\over 3}n_f
        \end{array}
      \right).
\end{equation}
The RG improved NLO cross sections are
defined as
\begin{equation}
\sigma_{NLO,RG}=\sigma_{NLO}+(\sigma_{RG}-\sigma_{RG}|_{NLO}),
\end{equation}where $\sigma_{RG}$ is the LO cross section calculated with RG
improved Wilson coefficients and $\sigma_{RG}|_{NLO}$ is the
expansion of $\sigma_{RG}$ up to NLO in QCD, which has already been
included in $\sigma_{NLO}$. If only $c_1$ or $c_2$ is non-zero at
the NP scale $\Lambda$, then the numerical solution of the RG
equation that sums the LO logarithms are
\begin{equation}\label{eqrun}
c_i(\mu_R)=c_1(\Lambda)X_{ij}r^{\gamma_j},\quad {\rm or} \quad c_i(\mu_R)=c_2(\Lambda)Y_{ij}r^{\gamma_j},
\end{equation}where $r=\alpha_s(\mu_R)/\alpha_s(\Lambda)$. The eigenvalues of
the one-loop anomalous dimension matrix $\gamma$, after being
divided by the overall factor $-{\alpha_s\over 2\pi}$ and $\beta_0$,
are
\begin{equation}
\gamma_i=(-0.630,\,0.843,\,-0.487,\,0.342,\,0.275,\,-0.155),
\end{equation}
where $\beta_0=(11N-2n_f)/3$. Furthermore,
\begin{eqnarray}
X_{ij}&=&\left(\begin{array}{cccccc}
0.127&	0.012&	0.206&	0.294&	0.292&	0.068\\
0.462&	-0.060&	0.577&	-0.577&	-0.462&	0.060\\
-0.130&	0.045&	0&	0&	-0.248&	0.332\\
0.470&	0.218&	0&	0&	-0.392&	-0.296\\
0.127&	0.012&	-0.206&	-0.294&	0.292&	0.068\\
0.462&	-0.060&	-0.577&	0.577&	-0.462&	0.060\\
       \end{array}\right),\nonumber\\ \nonumber\\
Y_{ij}&=&\left(\begin{array}{cccccc}
0.077&	-0.018&	0.105&	-0.105&	-0.076&	0.016\\
0.279&	0.087&	0.294&	0.206&	0.119&	0.014\\
-0.078&	-0.065&	0&	0&	0.064&	0.079\\
0.284&	-0.315&	0&	0&	0.101&	-0.071\\
0.077&	-0.018&	-0.105&	0.105&	-0.076&	0.016\\
0.279&	0.087&	-0.294&	-0.206&	0.119&	0.014\\
       \end{array}\right).
\end{eqnarray}
In Fig.\,~\ref{f2} we compare fixed order (NLO) and RG running of
the Wilson coefficients $c_1$, $c_2$ and $c_4$ for $\Lambda=30\,{\rm
TeV}$ assuming only $c_1$ (left panel) or $c_2$ (right panel) is
non-zero at the NP scale $\Lambda$, with the renormalization scale
ranges between $1\,{\rm TeV}$ and $30\,{\rm TeV}$. The running of
the Wilson coefficients at the NLO can be obtained from the
expansion of RG running in Eq.~(\ref{eqrun}).

\section{Applications to quark compositeness search at the LHC}

In this section we apply our NLO QCD results to the quark
compositeness search at the LHC ($\sqrt s=7\,{\rm TeV}$) through
dijet angular distribution measurement. For the numerical results
here we assume only $c_1(\Lambda)$ and $c_2(\Lambda)$ to be
non-zero, and parameterize them as
$c_{1(2)}(\Lambda)=4\pi\lambda_{1(2)}$. Following conventions used
in experimental analysis, for color-singlet case we have
$|\lambda_{1(2)}|=1(0)$, while for color-octet case we have
$|\lambda_{1(2)}|=0(1)$. We use the anti-$k_t$ jet
algorithm~\cite{Cacciari:2008gp} with energy recombination
scheme~\cite{Bayatian:2006zz} and the distance parameter $D=0.5$. To
be considered as one of the two leading jets, a jet is required to
satisfy the rapidity cut $|y|<3$. Moreover, we apply additional
constraints on jet rapidity as below
\begin{equation}\label{dijrap}
|y_b|=|y_1+y_2|/2<1.11, \,\,\chi=\exp{(|y_1-y_2|)}<16,
\end{equation}
to be consistent with
the CMS measurement~\cite{Khachatryan:2011as}. $\chi$ is
chosen as the dijet angular observable since after the Jacobian
transformation the t-channel dominant SM QCD dijet production is
almost flat on $\chi$. For massless jets, the invariant mass of
dijet can be expressed as
\begin{equation}\label{dijmass}
m_{jj}=\sqrt{p_{T1}p_{T2}}\sqrt{\chi+1/\chi-2\cos(\Delta \phi)},
\end{equation}where $\Delta \phi$ is the azimuthal angle between the two jets.
Considering the current experimental limits on the compositeness scale
as well as the large SM QCD dijet background, NP contributions can
only provide observable effects on the angular distribution in a
very high dijet invariant mass region. Thus, below we
will consider an invariant mass region between
 $[2\,{\rm TeV}$ and $3\,{\rm
TeV}]$,  and the part of theory parameter space with $\Lambda$
greater than $5\,{\rm TeV}$, for simplicity. In our numerical
calculations, we use
 CTEQ6.6 parton distribution functions \cite{Nadolsky:2008zw}
 and the corresponding running QCD
coupling constant.
Renormalization and factorization scales are set to be the average
transverse momentum of the two leading jets, unless otherwise
specified.

\subsection{LO results and analysis}\label{sec1}

\begin{figure}[h]\centering
\includegraphics[width=0.4\textwidth]{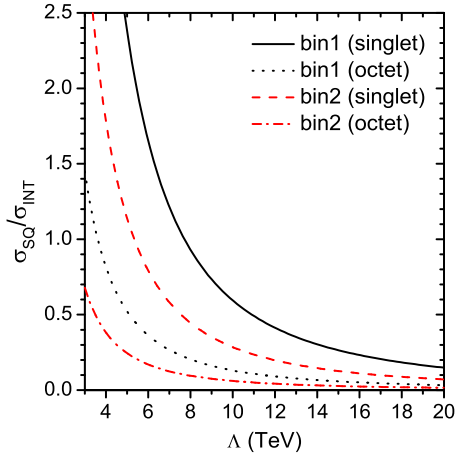}
\caption[]{Ratios of NP squared contributions ($\sigma_{SQ}$) to SM
and NP interference contributions ($\sigma_{INT}$) at the LO.}
\label{loratio}
\end{figure}

The LO total cross sections for dijet production induced by contact
interactions consists of interference contributions with SM QCD
amplitudes (denoted as $\sigma_{INT}$), as well as the NP squared
contributions (denoted as $\sigma_{SQ}$). Assuming that only $c_1$
and $c_2$ are non-zero, the additional LO contribution
\begin{eqnarray}
\sigma_{{\rm LO}} &=& \sigma_{INT} + \sigma_{SQ}  \\
&=&(\lambda_1b_{{\rm L},1}+\lambda_2b_{{\rm L},2})/\Lambda^2
+(\lambda_1^2b_{{\rm L},11}+\lambda_2^2b_{{\rm
L},22}+\lambda_1\lambda_2b_{{\rm L},12})/\Lambda^4,\nonumber
\end{eqnarray}
where $b_{{\rm L},i(ij)}$ are independent of
NP scale $\Lambda$ and $c_i=4 \pi \lambda_i$.
Instead of calculating the differential cross
sections with respect to $\chi$, we choose two representative bins
to investigate the influence of NP contributions to the dijet
angular distribution, which are $\chi=[1,6]$ (bin 1) and
$\chi=[6,11]$ (bin 2). The numerical results of $b_{{\rm
L},i(ij)}$ are shown in Table~\ref{locoe}. It can be seen that
absolute values of $b$ for bin 1 are much larger
than those for bin 2, which means the NP contributions could
change the shape of $\chi$ distribution significantly since the SM QCD
contributions are almost flat in $\chi$ distribution.
It is also indicated in Table~\ref{locoe} that, even for
a scale of $\Lambda$ as large as $5\,{\rm TeV}$, the NP squared
terms can still be comparable to or even larger than the
interference terms, especially for bin 1 of color-singlet case.
 Thus, the cross sections are not monotonous decreasing as
$\Lambda$ increases, and zero point occurs for certain $\Lambda$
value with destructive interference. For simplicity, we will focus
on the parameter region where the NP squared terms are relatively
small, i.e., $\sigma_{SQ}/\sigma_{INT}<1$. Fig.~\ref{loratio} shows
the absolute ratios of the NP contributions from the squared terms to
the ones from the interference as functions of $\Lambda$, where for the
color-singlet (octet) case we set $|\lambda_1|=1\,(0)$ and
$|\lambda_2|=0\,(1)$ following the standard convention as mentioned
before. Using the above condition, constraints on $\Lambda$ would be
$\gtrsim 8\,{\rm TeV}$ for the color-singlet case and $\gtrsim
4\,{\rm TeV}$ for the color-octet case, which will be applied
hereafter to our analysis.
\begin{table}[h!]
  \begin{center}
      \begin{tabular}{c|ccccc}
        \hline \hline
        [fb$\cdot$$(5\,{\rm TeV})^{2(4)}$]& $b_{{\rm L},1}$ & $b_{{\rm L},2}$
        &$b_{{\rm L},11}$&$b_{{\rm L},22}$& $b_{{\rm L},12}$
        \\ \hline
        bin 1 & -258 & -179 &614&93.4&259
        \\ \hline
        bin 2 & -99.1 & -70.4 &113&17.2&46.8
        \\ \hline \hline
      \end{tabular}
  \end{center}
  \caption{\label{locoe}LO coefficients of NP contributions.}
\end{table}

\subsection{NLO results and analysis}\label{sec2}

At NLO, the total cross sections for dijet production induced by the
contact interactions can be expressed as
\begin{eqnarray}
\sigma_{{\rm NLO}}&=&\big(\lambda_1(b_{{\rm
N},1}+a_1r)+\lambda_2(b_{{\rm N},2}+a_2r)\big)/\Lambda^2
+\big(\lambda_1^2(b_{{\rm N},11}+a_{11}r)\\
&&+\lambda_2^2(b_{{\rm N},22}+a_{22}r)+\lambda_1\lambda_2(b_{{\rm
N},12}+a_{12}r)\big)/\Lambda^4, \nonumber
\end{eqnarray}where $b_{{\rm N},i(ij)}$ and $a_{i(ij)}$ are
independent of the NP scale $\Lambda$, and $r=\ln(\Lambda/p_0)$. We
choose the reference scale
\begin{equation}\label{pscale}
p_0=\langle{m}_{jj}\rangle/\sqrt{\langle\chi\rangle+1/\langle\chi\rangle+2},
\end{equation}
where $\langle x\rangle$ denotes taking the average
value of $x$ in the given bin of kinematic variables. Hence,
$p_0$ is $1.04\,(0.77)\,{\rm TeV}$ for bin 1\,(2).

\begin{figure}[h]\centering
\includegraphics[width=0.35\textwidth]{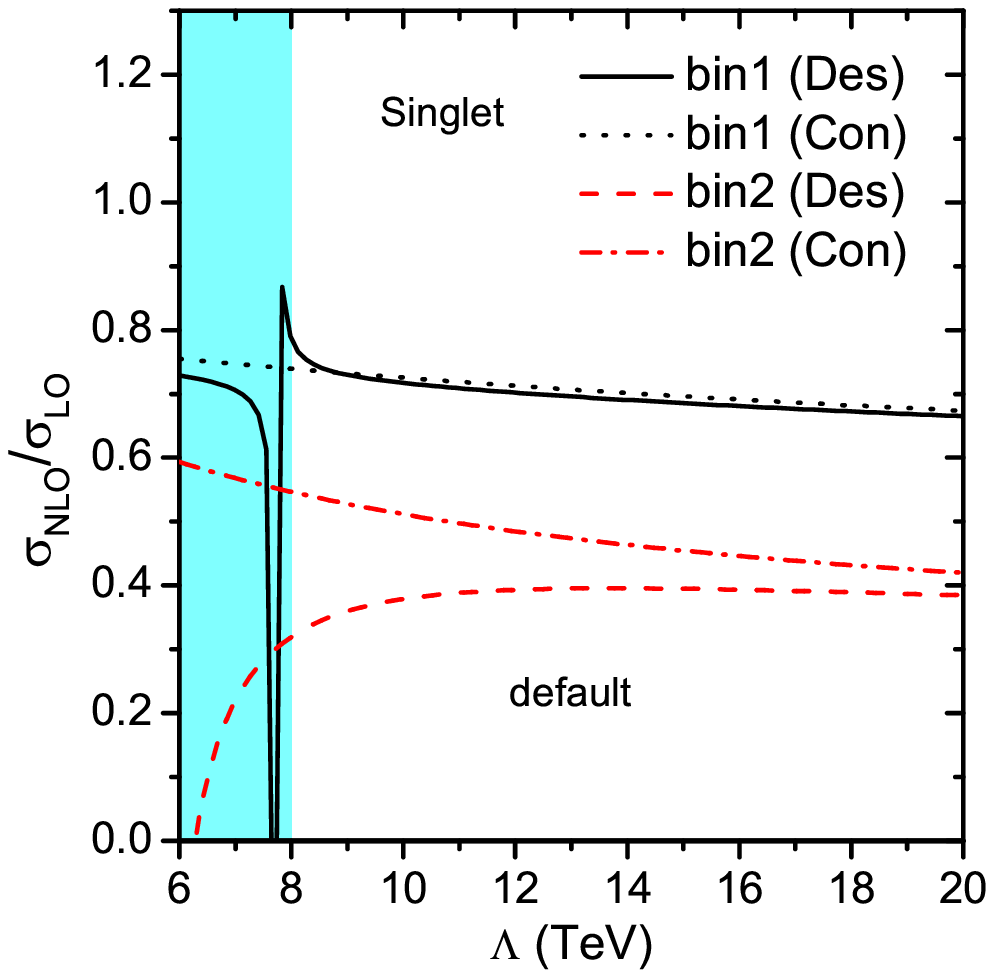}
\hspace{0.2in}
\includegraphics[width=0.35\textwidth]{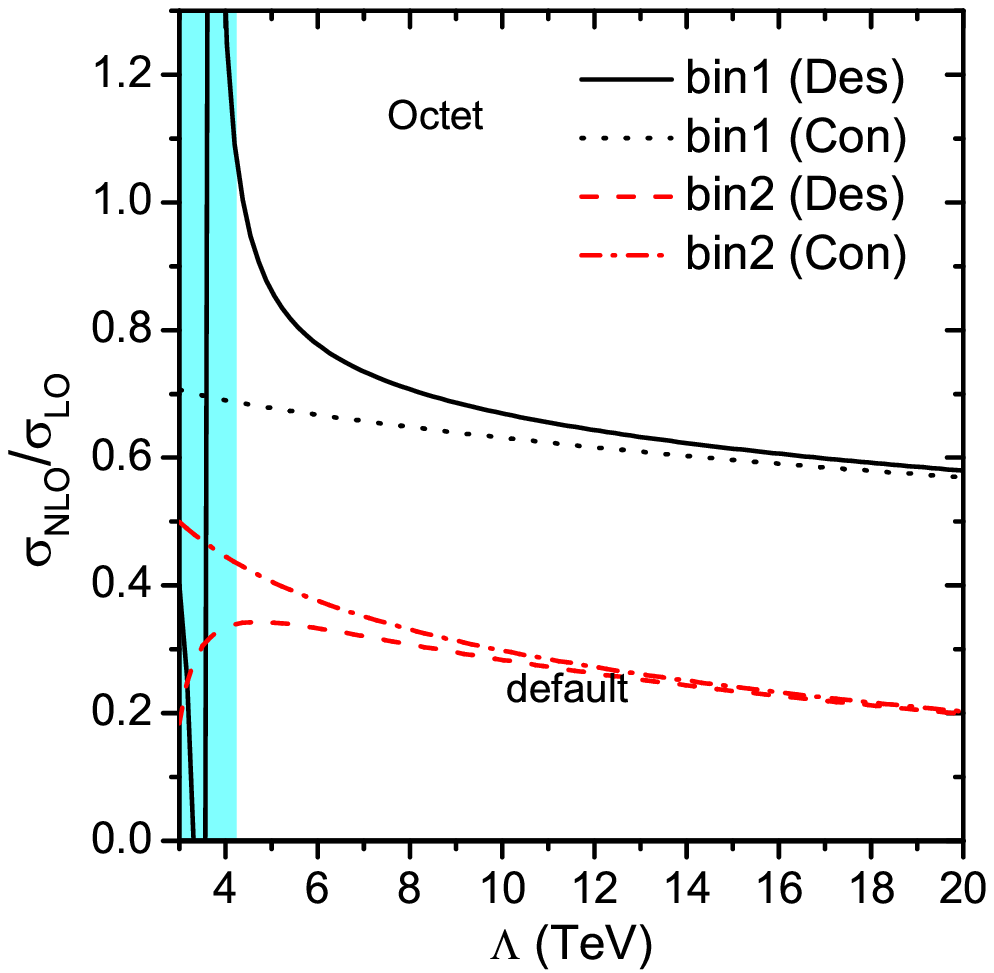}
\caption[]{NLO K-factors, as functions of $\Lambda$, using the
default definition of the LO cross sections and choice of scale.}
\label{nlok}
\end{figure}

\begin{figure}[h]\centering
\includegraphics[width=0.35\textwidth]{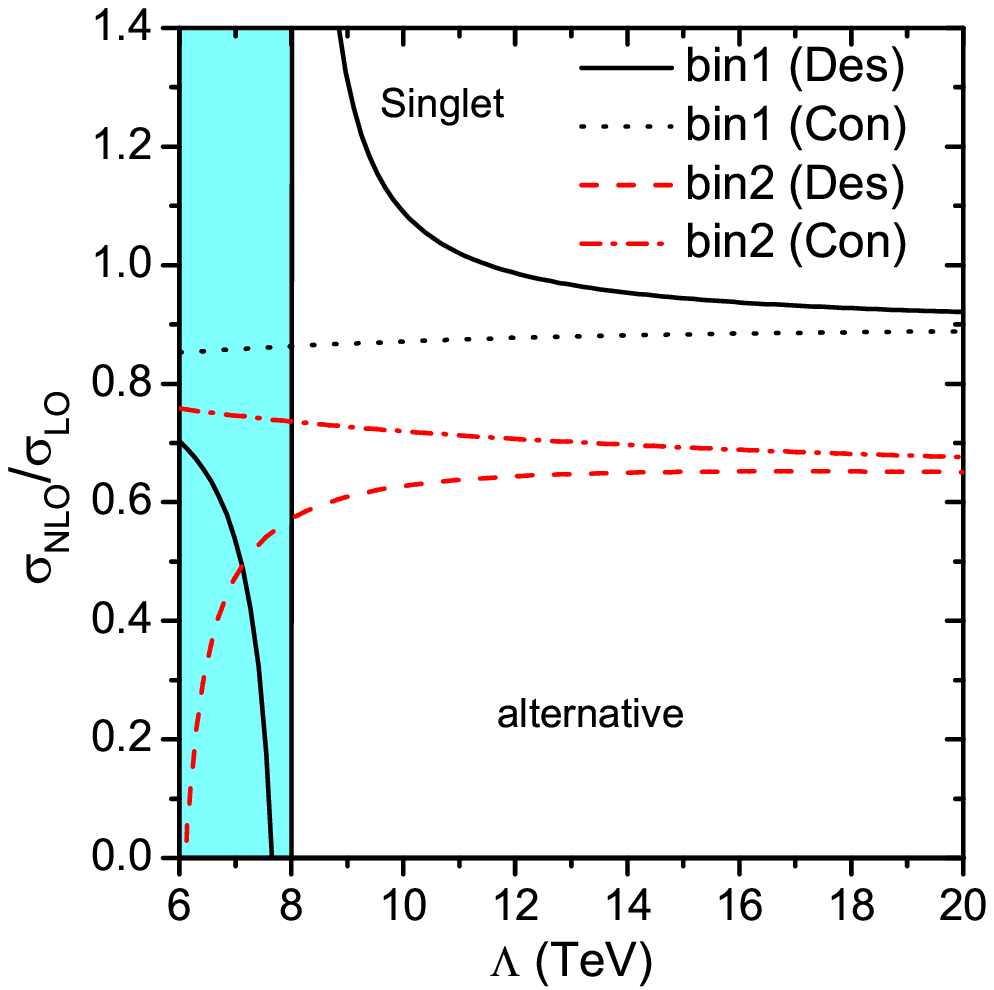}
\hspace{0.2in}
\includegraphics[width=0.35\textwidth]{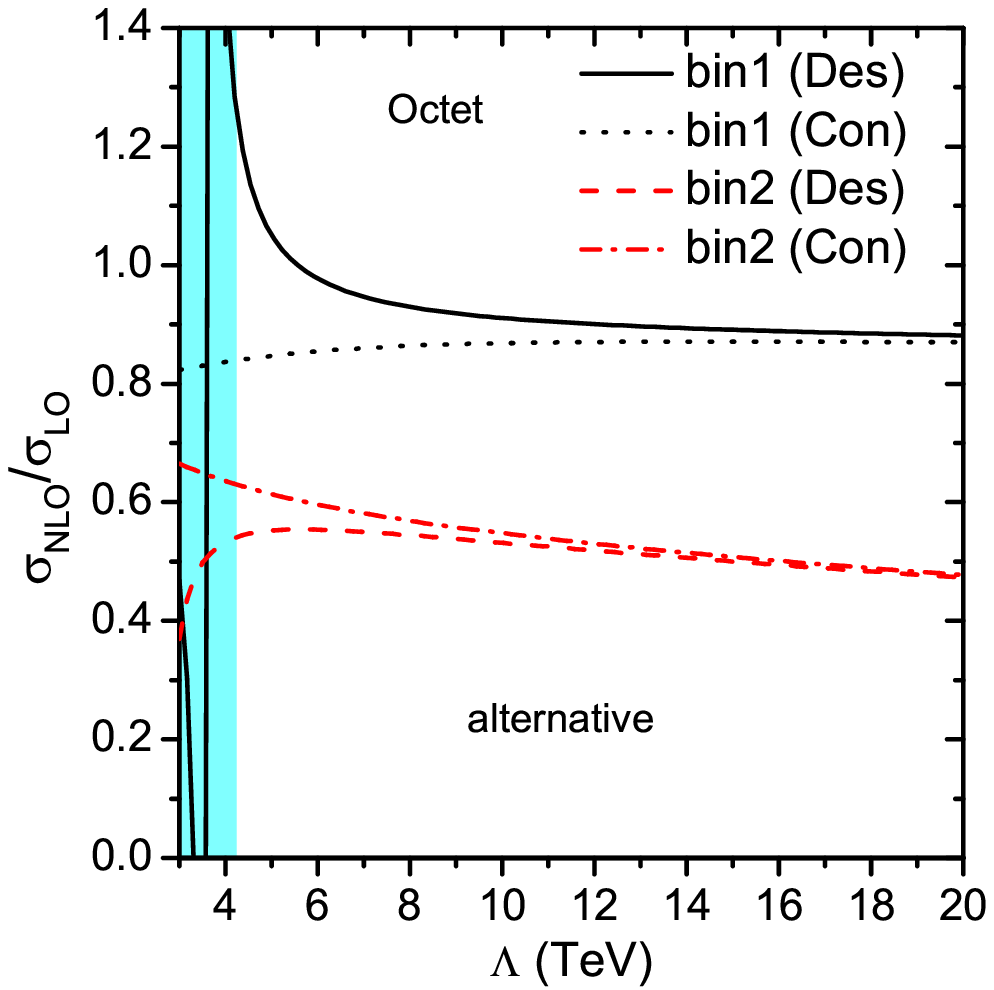}
\caption[]{NLO K-factors, as functions of $\Lambda$, using the
alternative definition of the LO cross sections and choice of
scale.} \label{nloknew}
\end{figure}

The contributions proportional to $r$ represent running effects of
the Wilson coefficients at NLO. Choosing same bins as in the LO
analysis, we get the results for $b_{{\rm N},i(ij)}$ and $a_{i(ij)}$
listed in Table~\ref{nlocoe}. Based on results from
Tables~\ref{locoe} and \ref{nlocoe}, we plot the NLO K-factors as
functions of the compositeness scale in Fig.~\ref{nlok} for both
cases with destructive and constructive interferences. The shadow
regions indicate parameter spaces where contributions from the NP
squared terms are larger than the ones from the interference terms.
K-factors are unstable there for destructive interference case due
to large cancelations of cross sections between NP squared
contributions and interference ones. Beyond the shadow regions, the
NLO QCD corrections reduce the absolute values of the NP
contributions to the cross sections significantly for all the
parameter cases. These are mainly due to the large constant terms
and also the large logarithms of $\Lambda$ from the virtual
corrections, cf. Eqs.~(3.4) and (3.5) in Sec.~\ref{asec1},
especially in the large $\chi$ region. From Fig.~\ref{nlok} we can
see that the K-factors deviate significantly from 1 especially for
bin 2. Thus one may doubt the reliability of the NLO results, i.e.,
the convergency of the perturbation series. Indeed, the small
K-factors are mainly due to the LO results used here. In order to
compare with the LO theoretical results used by the experimentalist,
by default we use fixed LO Wilson coefficients in the LO cross
sections here and below. If we use the NLO running Wilson
coefficients in the LO calculations instead, same as in the NLO
calculations, then the K-factors will increase due to the
suppression of the LO cross sections. Also the K-factors depend on
the QCD scale choices, especially for large $\chi$ region. If we set
the central scale to $p_{T,max}\exp(0.15|y_1-y_2|)$ as in the ATLAS
dijet study~\cite{Aad:2011fc}, where $p_{T,max}$ is the transverse
momentum of the hardest jet, then the K-factors will further
increase. In Fig.~\ref{nloknew} we show the K-factors calculated
using this alternative definition of the LO cross sections in the
perturbation series and choice of the scale. We can see that the NLO
results are well behaved with reasonable K-factors that are
relatively larger and stable as compared to the ones in
Fig.~\ref{nlok}.
\begin{table}[h!]
  \begin{center}
      \begin{tabular}{c|ccccc}
        \hline \hline
        [fb$\cdot$$(5\,{\rm TeV})^{2(4)}$]& $b_{{\rm N},1}$($a_1$)& $b_{{\rm N},2}$
        ($a_2$)&$b_{{\rm N},11}$($a_{11}$)&$b_{{\rm N},22}$($a_{22}$)& $b_{{\rm N},12}$($a_{12}$)
        \\ \hline
        bin 1 & -232(20) & -159(19) &506(-26)&74.3(-12)&172(-51)
        \\ \hline
        bin 2 & -68.3(8.7) & -44.1(9.2) &89.2(-4.9)&13.0(-2.3)&33.1(-9.8)
        \\ \hline \hline
      \end{tabular}
  \end{center}
  \caption{\label{nlocoe}NLO coefficients of NP contributions.}
\end{table}

\begin{figure}[h]\centering
\includegraphics[width=0.35\textwidth]{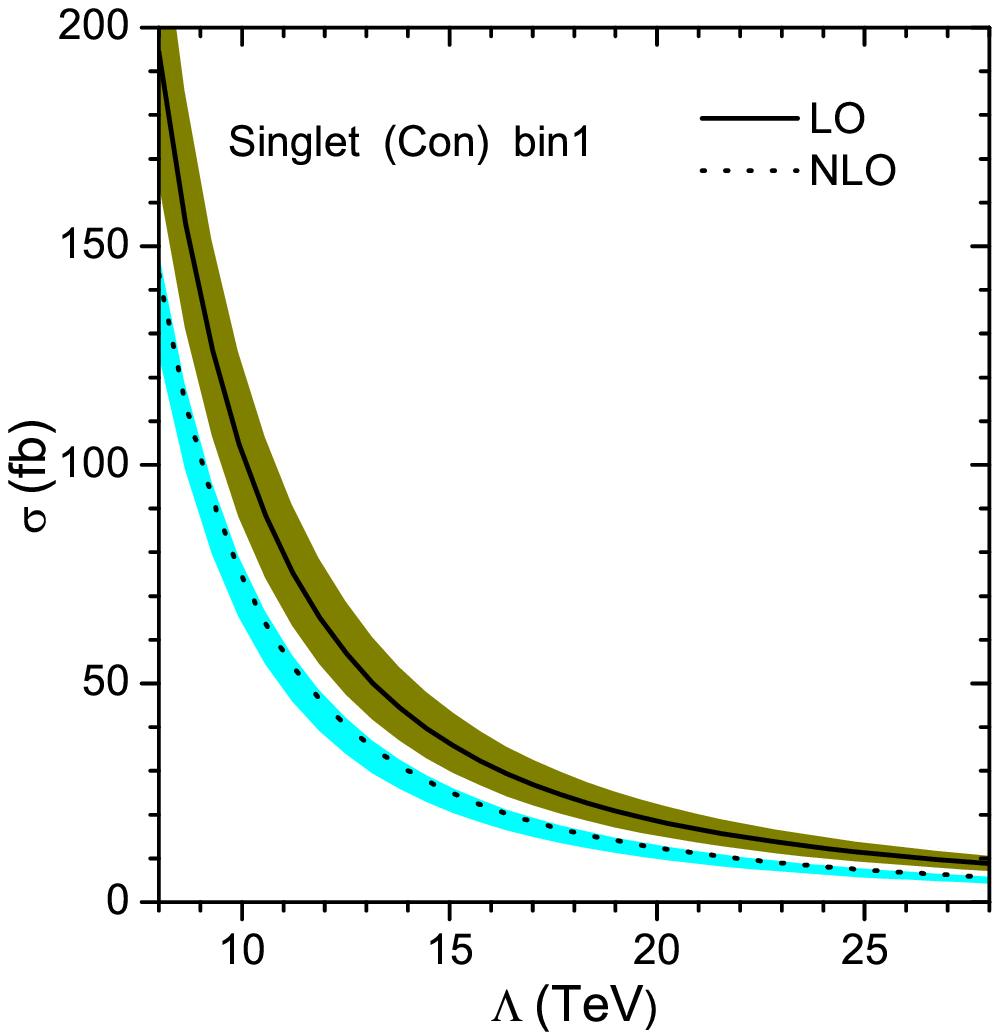}
\hspace{0.2in}
\includegraphics[width=0.35\textwidth]{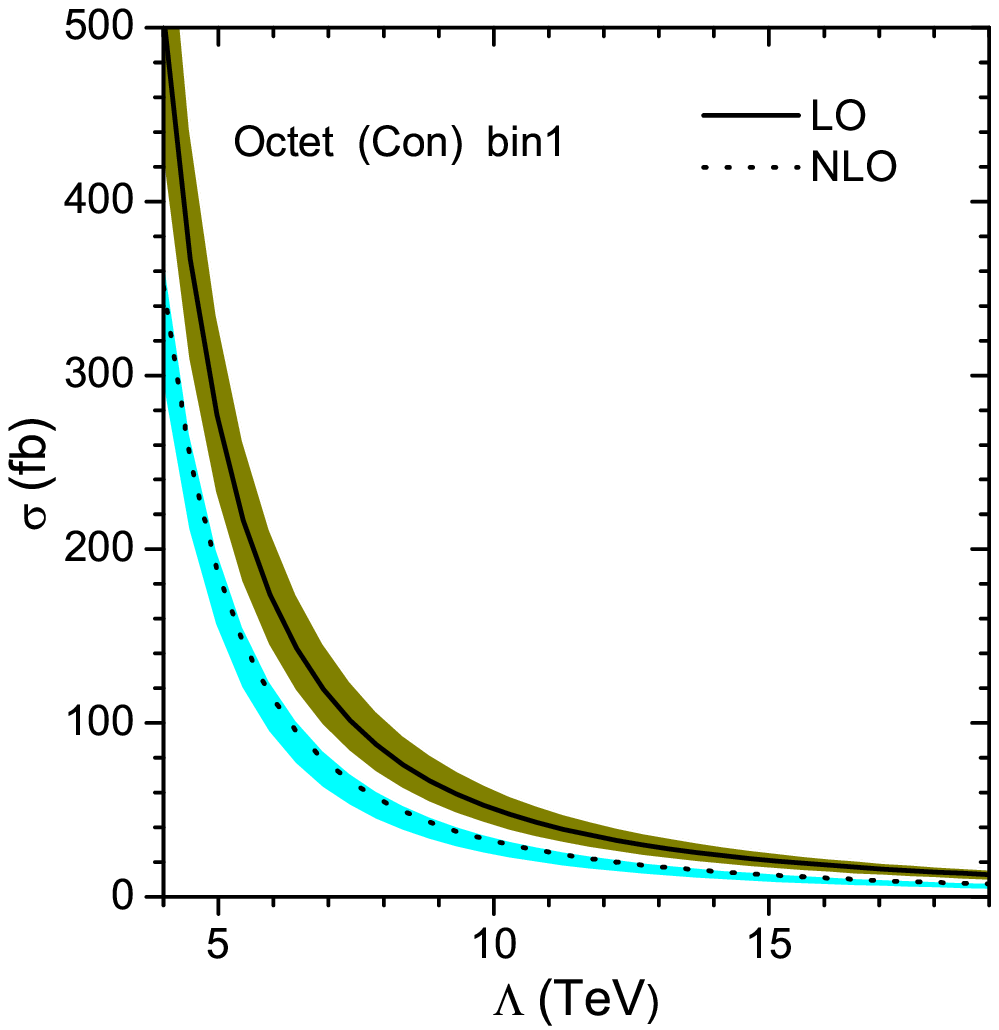}
\caption[]{LO and NLO total cross sections,
including scale uncertainties,
 as functions of $\Lambda$.} \label{scale}
\end{figure}

In Fig.~\ref{scale} we show NP contributions to the total cross
sections of bin 1 with scale uncertainties to further investigate
the improvement on scale dependence of the NLO results. The
uncertainties are calculated by varying the factorization and
renormalization scales, separately, for $\mu=\langle p_T\rangle/2$,
$\mu=\langle p_T\rangle$, and $\mu=2\langle p_T\rangle$, where
$\langle p_T\rangle$ is the average $p_T$ of the two leading jets.
Generally, we can see a reduction of the scale uncertainties for the NLO
results.

\subsection{RG improved NLO results}
\begin{figure}[h]\centering
\includegraphics[width=0.35\textwidth]{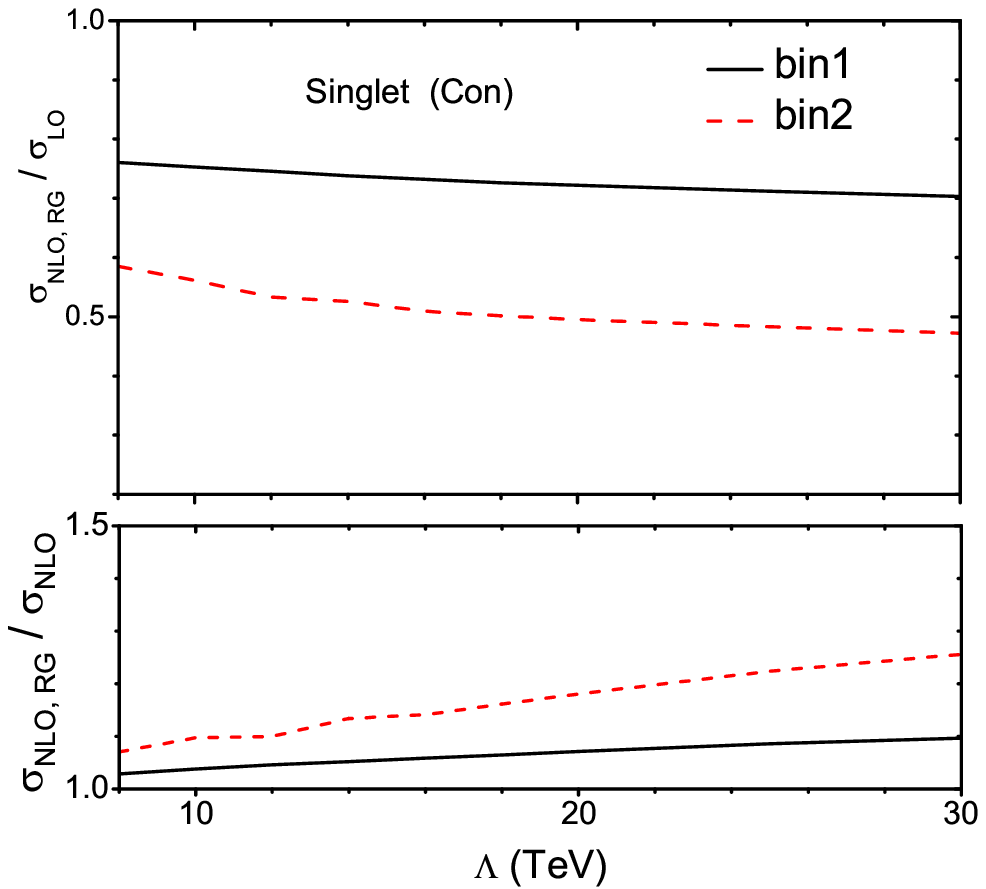}
\hspace{0.2in}
\includegraphics[width=0.35\textwidth]{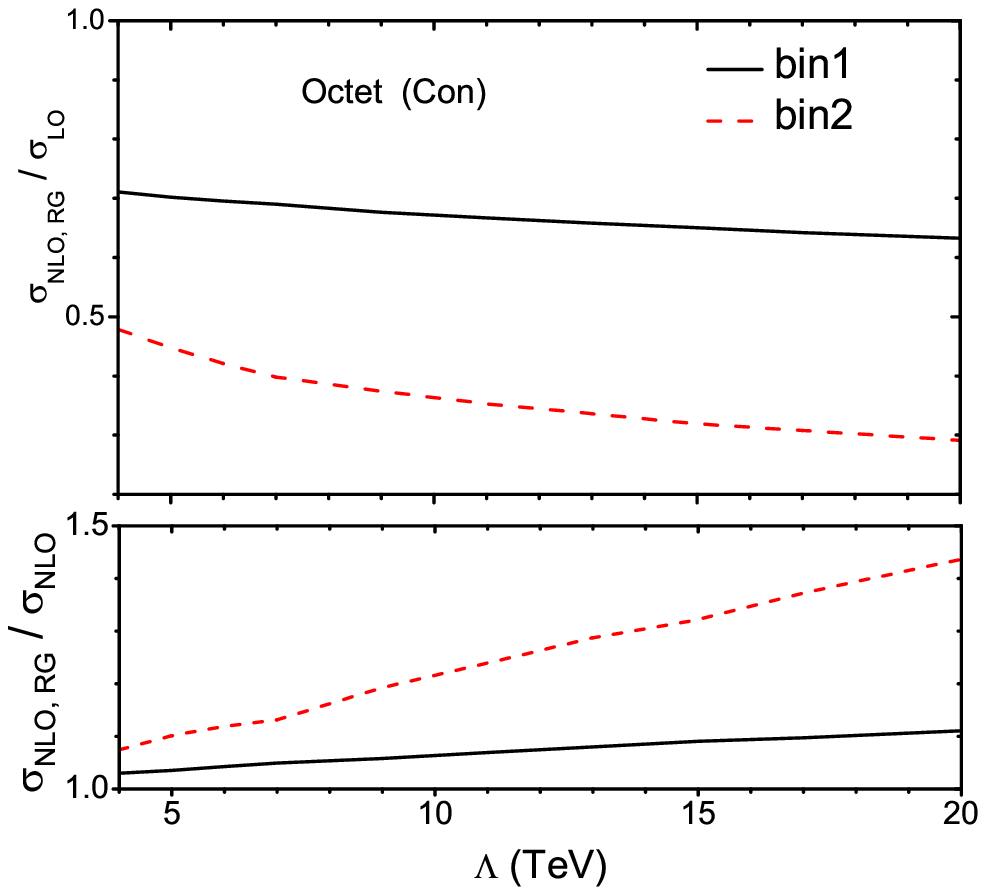}
\caption[]{K-factors of the RGE improved NLO cross sections, as
functions of $\Lambda$.} \label{rgk}
\end{figure}

As already mentioned in the analytical results, when the
compositeness scale is much higher than the typical energy scale
of experiments,
we can sum over large logarithms of $\Lambda$, induced from higher order
corrections, to improve the convergency of perturbative expansion.
This leads to the RG improved NLO cross sections $\sigma_{\rm RG,
NLO}$. It can deviate largely from the NLO cross sections in
the large $\chi$ region where the jet $p_T$ is
relatively small so that the running effects of Wilson coefficients are
important.
In Fig.~\ref{rgk},
We show the ratios $\sigma_{\rm
RG, NLO}/\sigma_{\rm LO}$ and $\sigma_{\rm RG, NLO}/\sigma_{\rm
NLO}$ as a function of $\Lambda$. In general, the higher
order corrections can increase the NLO total cross sections for both
color singlet and octet cases, and the amount depends on the
kinematic region considered. For example, for the bin 1
of color singlet case, the increase in $\sigma_{\rm RG, NLO}$ is about
3\% and 10\% of $\sigma_{\rm NLO}$
for $\Lambda=8\ {\rm TeV}$ and $30\ {\rm TeV}$, respectively.
Moreover, the
resumed contributions stabilize the K-factors of the total cross
sections as compared to the NLO results for high $\Lambda$ values, which
are around 0.7 and 0.5 for the bin 1 and the bin 2 of the color singlet case,
respectively.

\subsection{Exclusion limits of quark compositeness scale at the LHC}
\begin{figure}[h]\centering
\includegraphics[width=0.35\textwidth]{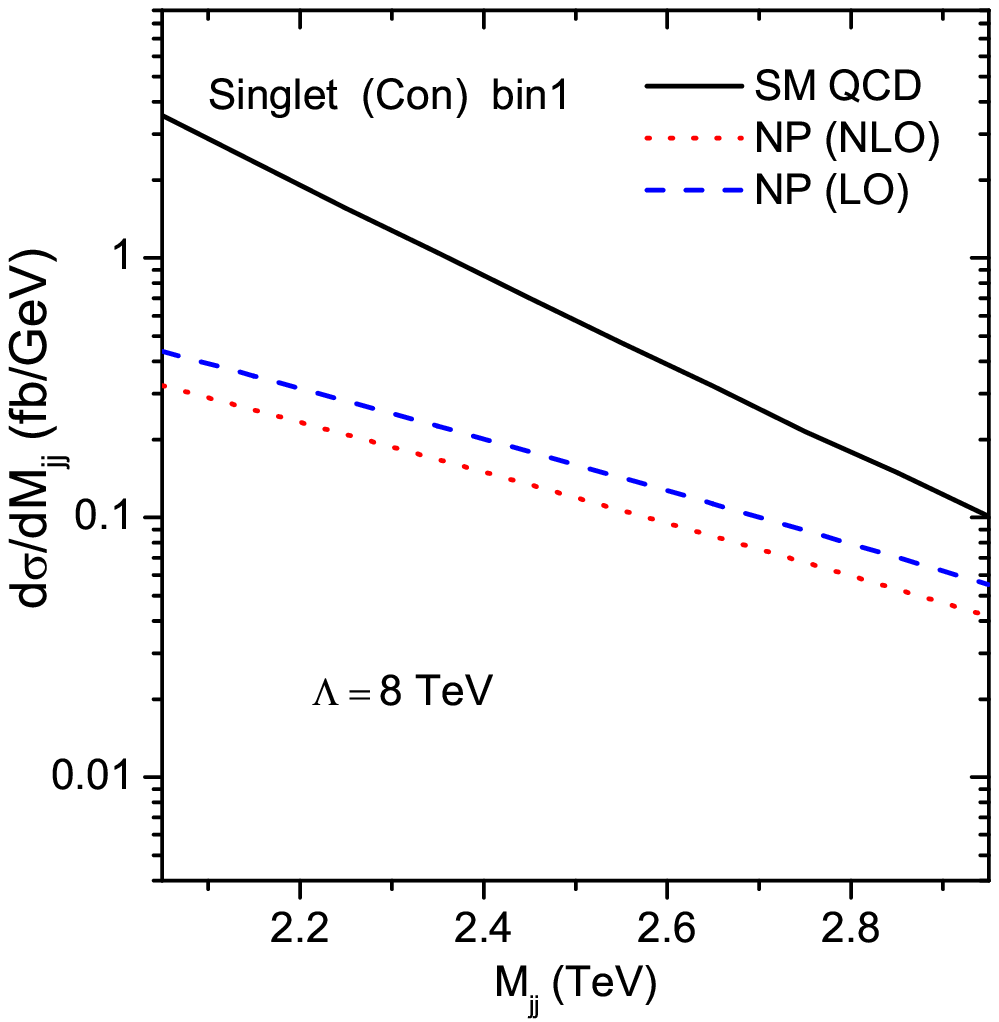}
\hspace{0.2in}
\includegraphics[width=0.35\textwidth]{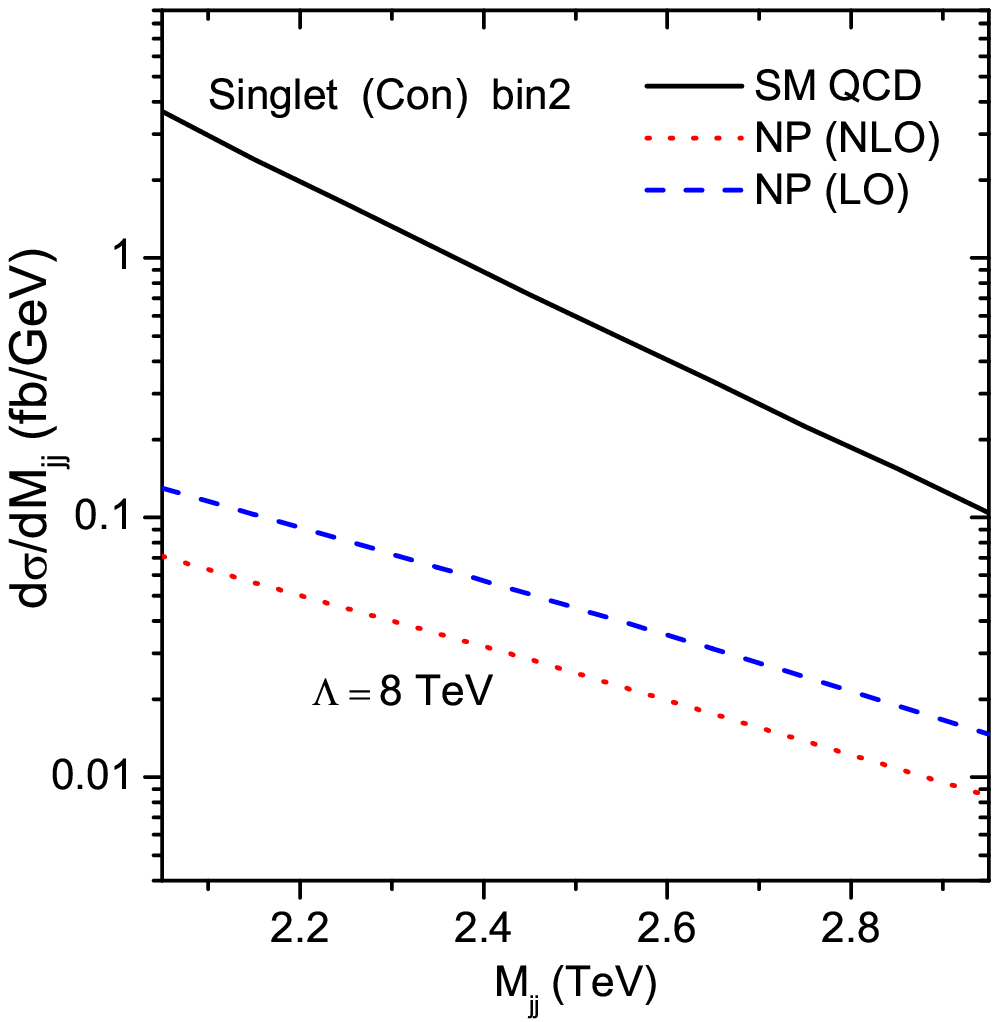}
\caption[]{Dijet invariant mass distributions from the pure SM QCD
contribution at NLO and the NP induced ones at both LO and NLO.}
\label{mbin}
\end{figure}
\begin{figure}[h]\centering
\includegraphics[width=0.4\textwidth]{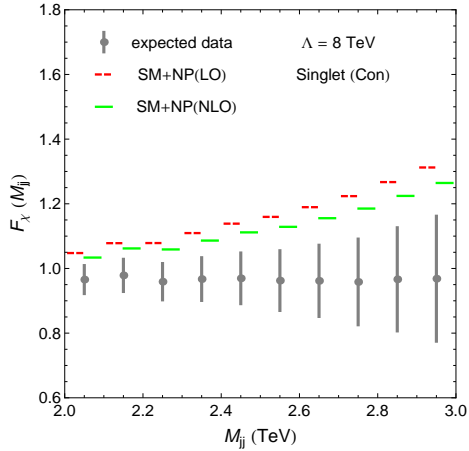}
\caption[]{$F_{\chi}(M_{jj})$ for pseudo data estimated by
pure SM QCD contribution, and for SM plus NP contributions
at the LO and NLO, respectively.} \label{ffbin}
\end{figure}

In order to directly compare our predictions to the experimental
measurements, we need to combine them with the pure SM QCD
contribution calculated at the NLO. We use a modified version of the
EKS code~\cite{Ellis:1992en} to calculate the SM QCD dijet
production at the NLO. We only consider the case of color-singlet
with constructive interference in this section, and we do not
include the RG improved corrections here since they are small for
the considered $\Lambda$ values. In Fig.~\ref{mbin}, we show
comparisons of differential cross section from NP and pure SM
contributions with invariant mass ranges from 2 to 3 $\rm TeV$ for
the color-singlet case with constructive interference and
$\Lambda=8\,{\rm TeV}$. As already mentioned before, the pure SM
contributions in the two bins are almost the same, while the NP
contributions are greatly different. To derive the expected
exclusion limits of the compositeness scale, we further divide the
invariant mass region $[2\,{\rm TeV},\,3\,{\rm TeV}]$ into 10 mass
bins with equal width, and define the measure in each mass bin,
$F_{\chi}(M_{jj})=\sigma_{\rm bin1}(M_{jj})/\sigma_{\rm
bin2}(M_{jj})$. In Fig.~\ref{ffbin}, we plot theoretical predictions
for $F_{\chi}(M_{jj})$ in different mass bins, where the (gray)
solid vertical line represents the pseudo data expected by pure SM
QCD contributions. The errors of the pseudo data include both the
estimated statistical and systematical errors. The former is
estimated by assuming Gaussian statistics with one standard
deviation. Here, we assume a large data set with
$\mathcal{L}=5\,{fb^{-1}}$ to calculate the statistical errors. We
note that most of the experimental systematic uncertainties cancel
in the ratio $F_{\chi}(M_{jj})$. For simplicity, we estimate an
overall experimental systematic uncertainty of 3\%, arising from the
jet energy calibration and jet $p_T$ resolution for all the mass
bins~\cite{Khachatryan:2011as}. The (colored) dashed and solid
horizonal lines represent the SM QCD contributions plus NP
contributions at LO and NLO, respectively. Fig.~\ref{ffbin} shows
that the NLO QCD corrections reduce the NP contributions to
$F_{\chi}(M_{jj})$.

\begin{figure}[h]\centering
\includegraphics[width=0.35\textwidth]{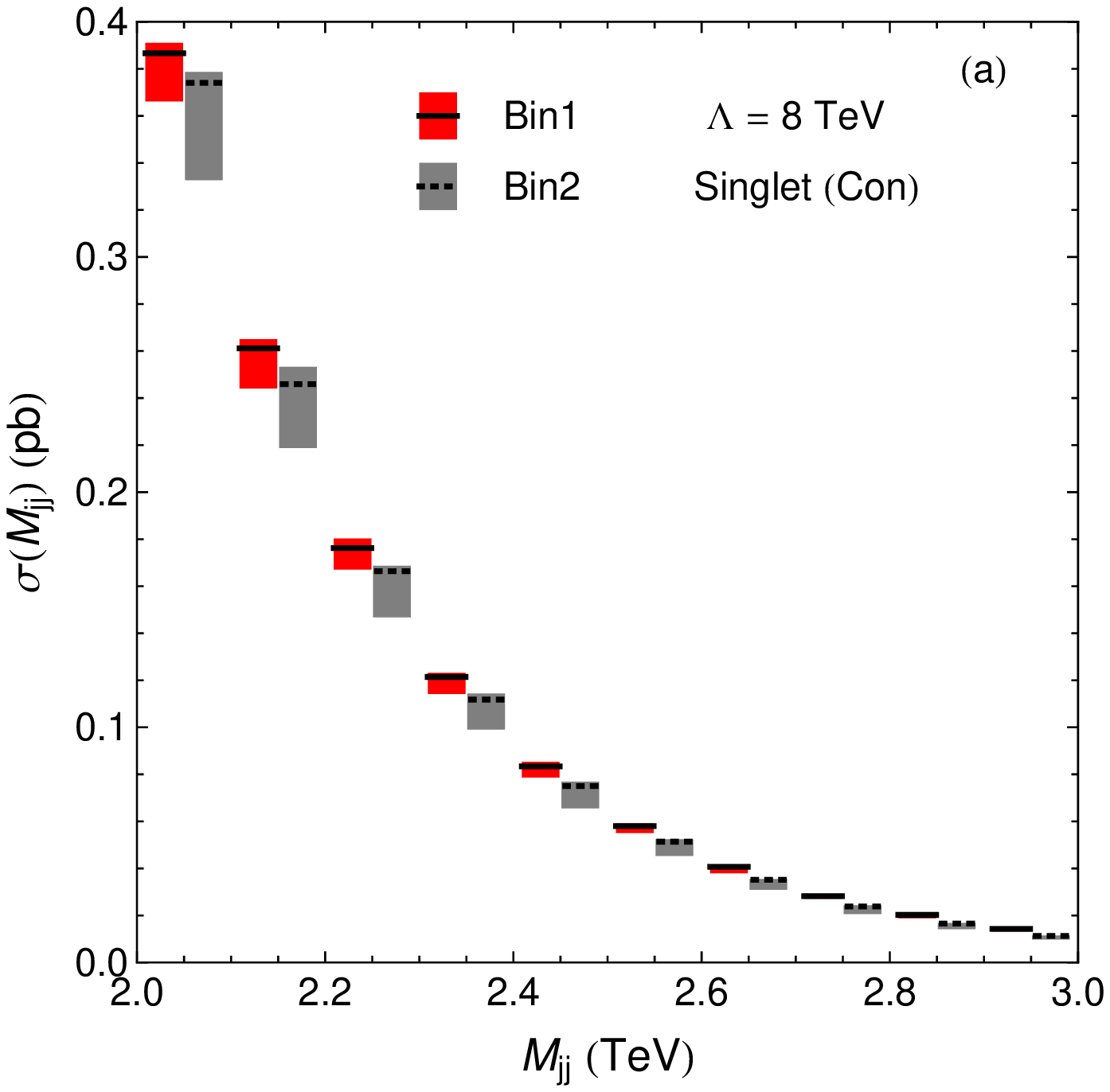}
\hspace{0.2in}
\includegraphics[width=0.35\textwidth]{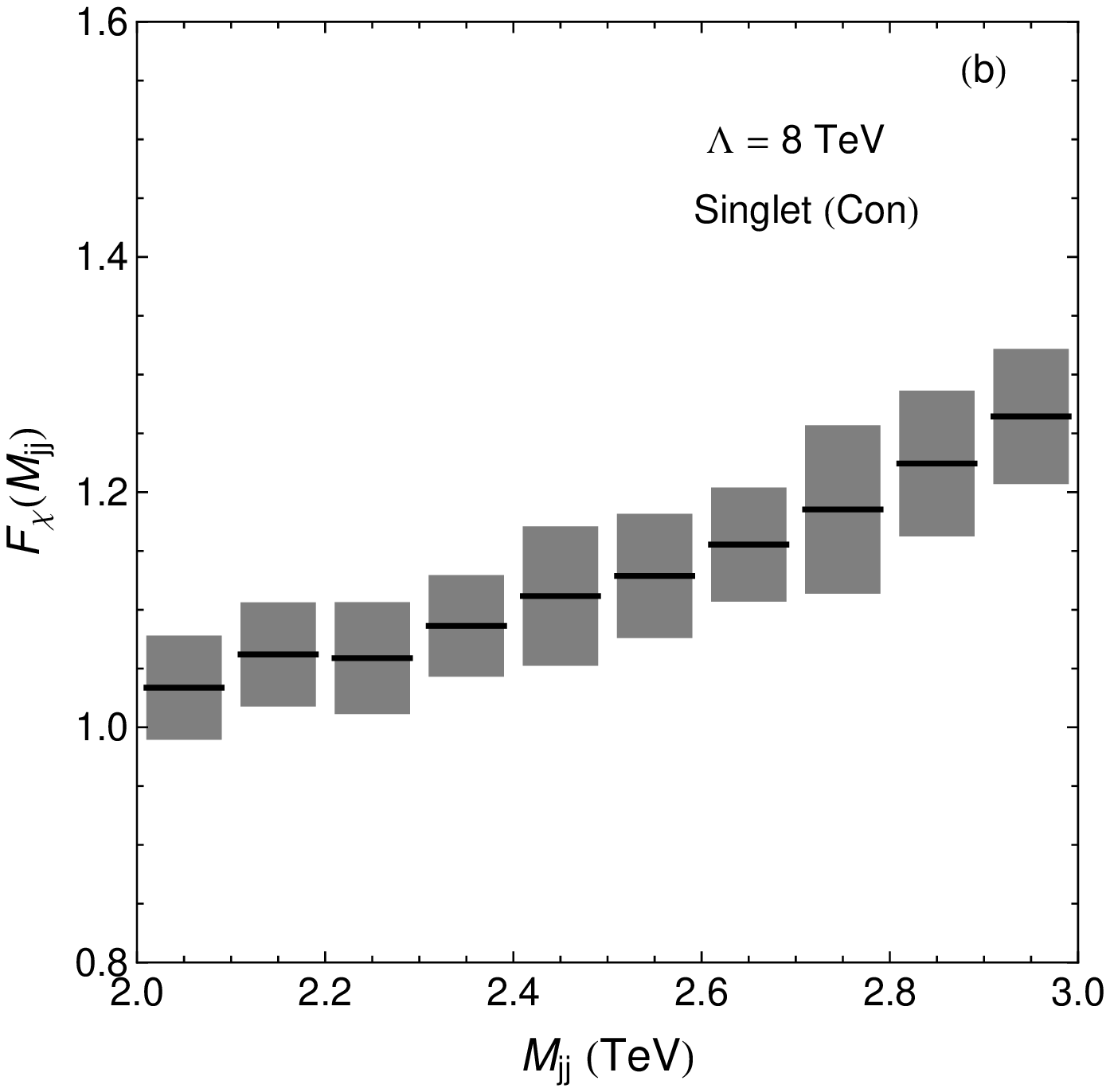}
\caption[]{(a) Total cross sections in bin 1 and bin 2 with scale
variations; (b) predictions of $F_{\chi}(M_{jj})$ with theoretical
errors included. Both of them are for SM plus NLO NP contributions.}
\label{uncer}
\end{figure}

On the other hand, the theoretical predictions also have
uncertainties due to parton distribution functions, non-perturbative
corrections, and most importantly the unknown higher order QCD
corrections. The first two are found to be small according to the
analysis in Ref~\cite{Khachatryan:2011as}. The conventional way to
estimate the uncertainty from unknown higher order corrections is to
examine its scale variations, i.e, to calculate the spread of the
cross section over a set of scale choices. In Fig.~\ref{uncer}(a),
we show scale variations of the cross sections in different bins for
SM plus NLO NP contributions, the spread are calculated by varying
the factorization and renormalization scales independently for
$\mu=\langle p_T\rangle/2$, $\mu=\langle p_T\rangle$, and
$\mu=2\langle p_T\rangle$. For $F_{\chi}(M_{jj})=\sigma_{\rm
bin1}(M_{jj})/\sigma_{\rm bin2}(M_{jj})$, we can independently vary
the scales in both the numerator and denominator factors. Since
there may be some correlations between these two parts in the
missing higher order corrections, it will lead to an overestimation
of the uncertainties. Here, we vary the scale simultaneously for
cross sections in bin 1 and bin 2, and take half of the total scale
variation of $F_{\chi}(M_{jj})$ as the estimated theoretical error.
The corresponding theoretical errors of $F_{\chi}(M_{jj})$ for SM
plus NLO NP contributions are shown in Fig.~\ref{uncer}(b), which
are around 4-7\% in different invariant mass bins.

\begin{figure}[h]\centering
\includegraphics[width=0.4\textwidth]{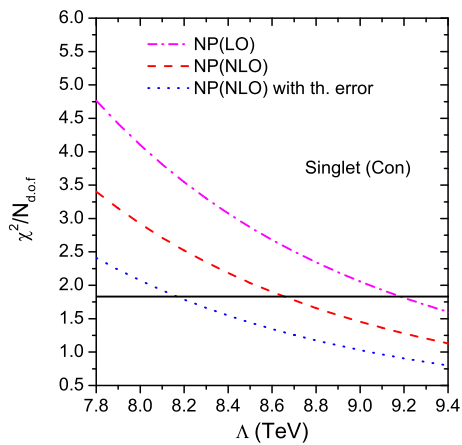}
\caption[]{$\chi^2/N_{d.o.f}$ as functions of the compositeness
scale $\Lambda$. Intersections of the curves with the horizontal
line show the exclusion limits at the 95\% C.L..} \label{limit}
\end{figure}

We perform a simple log-likelihood $\chi^2$ test for the hypothesis
of NP with
\begin{equation}
\chi^2=\sum_{i=1,\,10}
\frac{(F_{\chi}^{SM+NP}(i)-F_{\chi}^{SM}(i))^2}{\Delta_{exp}^2(i)+\Delta_{th}^2(i)},
\end{equation}where $F_{\chi}^{SM}(i)$ represents the pure SM contribution of
$F_{\chi}(M_{jj})$ in the $i$th mass bin, which we assume to be the
expected data, and $F_{\chi}^{SM+NP}(i)$ is the theory prediction
given by SM plus NP contributions. $\Delta_{exp,\,th}$ represents
the corresponding experimental errors and theoretical errors of
$F_{\chi}^{SM+NP}(i)$, and we do not consider possible correlations
of errors in different mass bins. The $\chi^2/N_{d.o.f}$ with
$N_{d.o.f}=10$ are shown in Fig.~\ref{limit} as functions of the
compositeness scale $\Lambda$ for 3 cases, i.e, SM plus LO or NLO NP
contributions without including the theoretical uncertainty, and SM
plus NLO NP contributions with theoretical uncertainty included.
Exclusion limits (95\% C.L.) of $\Lambda$ can be read directly from
Fig.~\ref{limit} as intersections of the curves with the horizontal
line. We can see that the theoretical uncertainty has a large effect
on the exclusion limit since they are comparable with the
experimental errors. With more data collected at the LHC, the
statistical error will further decrease, then theoretical
uncertainty will play a much more important role in the measurement.

\section{Conclusion}
In conclusion, we have calculated the exact NLO QCD corrections to
dijet production at the LHC, induced by quark contact
interactions with different color and chiral structures from new
physics. By applying our results to quark compositeness search at
the LHC, we show that the NLO QCD corrections can lower the NP cross
sections by several tens percent, depending on
the choice of the theory parameters and
the kinematic regions considered.
Moreover, the NLO QCD corrections reduce the
dependence of the cross sections on factorization and
renormalization scales. We also calculate the renormalization group
improved NLO cross sections by summing over the large logarithms
induced  from the
running of Wilson coefficients, which are found to stabilize the
K-factors at large quark compositeness scales. We further
investigate the NLO QCD effects on the corresponding experimental
observables and study the exclusion limits of the quark compositeness
scale.

\section*{Appendix}
In this appendix we briefly introduce our numerical code CIDIJET2.3,
developed for the NLO QCD calculation of the
dijet production induced by quark contact interaction.

Basically, this code is calculating the double differential cross
sections in terms of the dijet invariant mass $M_{jj}$ and the angle
parameter $\chi$. There are two calculation modes in this code. The
first one is to numerically evaluate the fixed order double
differential cross sections, in a kinematic bin specified by its
range of $M_{jj}$ and $\chi$, for various values of $\Lambda$ and
$c_i$ (or $\lambda_i$). For $c_{i}=4\pi \lambda_{i}$, the fixed
order (LO or NLO) cross section can be written as
\begin{eqnarray}
\sigma_{bin}&=&\sum_{i=1}^6(\lambda_i(b_{i}+a_ir))/\Lambda^2
+\sum_{i=1}^6(\lambda_i^2(b_{ii}+a_{ii}r))/\Lambda^4 \nonumber\\
&&+\sum_{i=1,3,5}(\lambda_i\lambda_{i+1}(b_{ii+1}+a_{ii+1}r))/\Lambda^4\nonumber\\
&&+\sum_{i=1,2,5,6}(\lambda_i\lambda_{4}(b_{i4}+a_{i4}r))/\Lambda^4.
\end{eqnarray}
with $r=\ln(\Lambda/p_0)$, and $p_0$ is defined as in
Eq.~(\ref{pscale}). The code can calculate all the above
coefficients ($a$'s and $b$'s) directly, thus users do not need to
repeat the calculations for different $\Lambda$ and $\lambda_i$
values. Since QCD interaction preserves parity conservation, we
should have $b(a)_{1,2}=b(a)_{5,6}$,
$b(a)_{11,22,12}=b(a)_{55,66,56}$, and $b(a)_{14,24}=b(a)_{54,64}$.
The code has two dynamic QCD scale choices, which is average $p_T$
of two leading jets used by the CMS collaboration and
$p_{T,max}\exp(0.15|y_1-y_2|)$ used by the ATLAS collaboration, and
provides inputs of pre-factors for both renormalization and
factorization scale to study the scale variations. Another mode is
for the calculation of higher order corrections in addition to the
NLO cross sections, which arise from the renormalization group
running of the Wilson coefficients, as discussed in
section~\ref{rgc}. They are only significant at very large $\Lambda$
values and need to be recalculated for different inputs of $\Lambda$
and $\lambda_i$. The CIDIJET2.3 code is publicly available by
request.

\begin{acknowledgments}
This work was supported by the U.S. DOE Early Career Research Award
DE-SC0003870 by Lightner-Sams Foundation, and the National Natural
Science Foundation of China, under Grants No.11021092 and
No.10975004. C.P.Y acknowledges the support of the U.S. National
Science Foundation under Grand No. PHY-0855561. We appreciate
helpful comments and communications with Sunghoon Jung, P. Ko,
Yeo Woong Yoon, and Chaehyun Yu.
\end{acknowledgments}

\end{document}